\documentclass[aps,prl,twocolumn,epsfig,amsmath,amssymb]{revtex4}
\usepackage[english]{babel} \usepackage{latexsym}
\usepackage{graphicx} \usepackage{subfigure} \usepackage{epsfig}
\usepackage{psfrag}

\begin{document}

\title{Two-dimensional bright solitons in dipolar Bose-Einstein condensates} 
\author{P. Pedri and L. Santos} 
\affiliation{
\mbox{Institut f\"ur Theoretische Physik III, Universit\"at
Stuttgart, Pfaffenwaldring 57, D-70550, Stuttgart, Germany}\\
}

\begin{abstract}  
%
%

We analyze the physics of bright solitons in two-dimensional 
dipolar Bose-Einstein condensates. These solitons are not possible in 
short-range interacting gases. In particular, we discuss 
the necessary conditions for the existence of stable 
2D bright solitary waves. The stability of the solutions is 
studied by means of the analysis of lowest-lying excitations. 
Additionally, we study the 
scattering of solitary waves, which contrary to the contact-interacting 
case, is inelastic, and could lead to fusion of solitary waves. 
Finally, the experimental possibilities for observability are discussed.
\end{abstract}  
\pacs{03.75.Fi,05.30.Jp} \maketitle


During the last years, the physics of ultracold atomic and molecular gases 
has attracted a large interest. Although these gases are very dilute, their 
properties are crucially determined by the interparticle interactions \cite{Review}.   
Up to now, only short-range van der Waals interactions have played 
a significant role in typical experiments. However, very recent developments 
are paving the way towards a new fascinating research area, namely that 
of degenerate dipolar gases. A major breakthrough has been very recently 
performed at Stuttgart University, where a Bose-Einstein condensate (BEC) of Chromium 
atoms has been realized \cite{Chromium}. 
Chromium atoms are particularly interesting, since they present 
a large magnetic dipole moment, $\mu=6\mu_B$. Hence, Chromium BEC constitutes the 
first realization of a degenerate dipolar gas. 
On the other hand, recent developments on cooling and trapping of molecules \cite{Molec1},  
on photoassociation experiments \cite{PA}, and on Feshbach resonances of binary mixtures 
\cite{Feshbach}, open exciting perspectives towards the realization of a 
degenerate gas of polar molecules. Particularly interesting in this sense is the recent 
creation of RbCs molecules in their absolute vibronic ground state $X^1\Sigma^+$, with a calculated 
electric dipole moment of $1.3$ D \cite{DeMille}. 

In addition to the usual short-range forces, dipolar particles 
oriented by some external electric or magnetic field interact via 
dipole-dipole interaction, which is long-range and 
anisotropic, being partially attractive. New exciting physics is 
therefore expected in these systems. In particular, recent theoretical 
analyses have shown that the stability and excitations of dipolar gases are crucially  
determined by the trap geometry \cite{Stability,Excitations,Roton}. 
Ultracold dipolar particles are also attractive in the context of strongly-correlated atomic gases 
\cite{DipLat,FQHE}, as physical implementation of 
quantum computation ideas \cite{QInf}, and for the study of ultracold chemistry \cite{Chemistry}.

The nonlinearity of the BEC physics is one of the major consequences 
of the interparticle interactions. In this sense, resemblances 
between BEC physics and nonlinear physics (in particular nonlinear optics) 
have been analyzed in detail. Several remarkable experiments have been 
reported in this context, including four-wave mixing \cite{4WM}, BEC 
collapse \cite{Collapse}, and 
the creation of dark, bright and gap solitons \cite{Solitons}. 
The physics of BEC solitons has indeed arouse 
a large interest. For short-range interactions and at sufficiently low temperature, 
the BEC physics is provided by a nonlinear Schr\"odinger 
equation (NLSE) with cubic nonlinearity (Gross-Pitaevskii equation) \cite{Review}. 
In 1D this equation admits  
solitonic solutions for bright (dark) solitons \cite{Zakharov} 
for attractive (repulsive) interatomic interactions, the equivalent of 
self-focusing (self-defocusing) nonlinearity in Kerr media. 

The equation determining the physics of a dipolar BEC is, as we  
discuss below, a NLSE with nonlocal nonlinearity, 
induced by the dipole-dipole interaction. Some interesting effects 
related with nonlocal nonlinearity 
have been studied in different contexts. In particular, 
it has been suggested that the nonlocality may play an important role 
in the physics of solitons and modulation instability \cite{Kivsharpapers}. 
Recent experiments in nematic liquid crystals have confirmed 
some of these predictions \cite{Nematics}. Particularly interesting is the 
possibility of stabilization of localized waves in cubic nonlinear materials with a symmetric 
nonlocal nonlinear response \cite{Bang}.

In this letter we are interested in the physics of solitons in 
two-dimensional dipolar condensates. Whereas for short-range interacting BEC 
a 2D stable solitary wave is not possible, we show that 
the dipole-dipole interactions may stabilize a 2D solitary wave. 
We discuss the appropriate conditions at which this is 
possible, which involve the tuning of the dipole-dipole interactions as 
discussed in Ref.~\cite{Tuning}. We study the stability of the solitary 
waves, discussing its lowest-lying excitations. The scattering of 2D solitary 
waves is also analyzed. We show that this scattering is inelastic, and 
it can lead under appropriate conditions to the fusion of solitary waves. 
Finally, we conclude with a discussion about the experimental 
feasibility and generation of the solitary waves.



In the following we consider a BEC of $N$ particles with electric dipole $d$
(the results are equally valid for magnetic dipoles). The dipoles are oriented 
by a sufficiently large external field, and hence interact via a dipole-dipole potential:
\begin{equation}
V_d(\vec r)=g_d \left (1-3\cos^2\theta\right )/r^3,
\label{Vd}
\end{equation}
where $g_d=\alpha Nd^2/4\pi\epsilon_0$, with 
$\epsilon_0$ the permittivity of vacuum, $\theta$ is the angle formed by the vector 
joining the interacting particles and the dipole direction, and $\alpha$ is a parameter 
tunable as follows. As in Ref.~\cite{Tuning} we consider that the electric 
field which orients the dipoles  
is a combination of a static field along the $z$-direction and a fast rotating field 
in the radial plane $\vec E(t)=E \left \{  
\cos\phi \hat z+\sin\phi [\cos (\Omega t) \hat x +\sin  (\Omega t)\hat y ]
\right\}$. The rotating frequency $\Omega$ is sufficiently slow to guarantee that
the dipole moments follow adiabatically the field $E(t)$, but sufficiently fast 
such that the atoms do not significantly move during the time $1/\Omega$ \cite{Tuning}. 
In this case, the interaction can be averaged over a rotation period, 
resulting in a tunable dipole-dipole potential, with $\alpha = (3\cos ^2\phi -1)/2$, which   
may range from $1$ to $-1/2$. The tunability of the 
dipole-dipole interaction will become crucial in our discussion of 
2D solitary waves.


The physics of the dipolar BEC at sufficiently low temperatures 
is provided by a NLSE with nonlocal nonlinearity \cite{Stability}:
\begin{eqnarray}
i\hbar\frac{\partial}{\partial t}\Psi(\vec r,t)&=&
\left [ 
-\frac{\hbar^2}{2m}\nabla^2+U(\vec r)
+g|\Psi(\vec r,t)|^2 \right\delimiter 0 \nonumber \\
&+& \left\delimiter 0  \int d\vec r' V_d(\vec r-\vec r')|\Psi(\vec r',t)|^2
\right ]\Psi(\vec r,t),
\label{GPE}
\end{eqnarray}
where $\int |\psi(\vec r,t)|^2 d{\vec r}=1$, and 
$g=4\pi\hbar^2aN/m$ is the coupling constant which characterizes 
the contact interaction, with $a$ the $s$-wave scattering length. 
In the following we consider $a>0$, i.e. repulsive short-range interactions. 
We assume an external trapping potential 
$U(\vec r)=m\omega_z^2 z^2/2$, with no trapping in the $xy$-plane.


In the following we are interested in the possibility to achieve a 2D 
localized wave in dipolar BECs. In order to have a good insight on this issue, 
we introduce a Gaussian Ansatz for the wavefunction:
\begin{equation}
\Psi_0(\vec r)=\frac{1}{\pi^{3/4} l_z^{3/2} L_\rho L_z^{1/2}}
\exp \left ( -\frac{x^2+y^2}{2 l_z^2 L_\rho^2}-\frac{z^2}{2l_z^2 L_z^2} \right ),
\end{equation}
where $l_z=\sqrt{\hbar/m\omega_z}$, and $L_\rho$ and $L_z$ are dimensionless variational 
parameters related with the widths in the $xy$-plane and the $z$-direction, respectively.
Using this Ansatz the energy of the system reads: 
\begin{equation}
\frac{2E}{\hbar\omega_z}= 
\frac{1}{L_\rho^2}+ \frac{1}{2L_z^2}+\frac{L_z^2}{2}+
\frac{1}{\sqrt{2\pi}L_\rho^2L_z}
\left [
\frac{\tilde g}{4\pi}+
\frac{\tilde g_d}{3}f\left ( \frac{L_\rho}{L_z}\right )
\right ]
\end{equation}
where $\tilde g=2g/\hbar\omega_zl_z^3=8\pi N a/l_z$ and  
$\tilde g_d=2g_d/\hbar\omega_zl_z^3$, and 
$
f(\kappa)=(\kappa^2-1)^{-1}\left [ 2\kappa^2+1-3\kappa^2 H(\kappa) \right ]$, 
with $H(\kappa)={\rm arctan}(\sqrt{\kappa^2-1})/\sqrt{\kappa^2-1})$. 

\begin{figure}[ht] 
\begin{center}
\includegraphics[width=7.0cm]{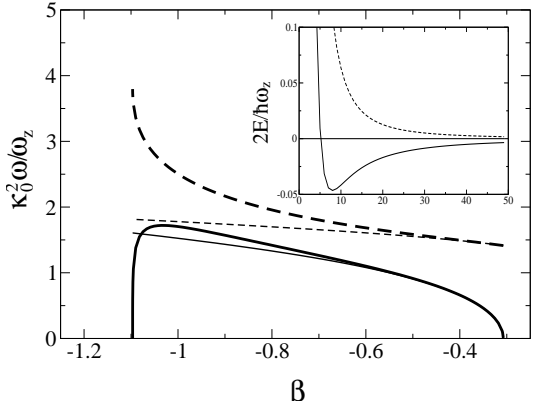}
\end{center} 
\vspace*{-0.2cm} 
\caption{Breathing (bold dashed) and $m=\pm 2$ quadrupole (bold solid) mode  
for $\tilde g=20$, with $\kappa_0=L\rho/L_z$. Results from a purely 2D calculation are shown in thin lines. 
Inset: $E(\kappa)$ for $\tilde g=500$, and $\beta=-0.10$ (dashed) and $\beta=-0.20$ (solid).}
\label{fig:1}  
\vspace*{-0.1cm}  
\end{figure}

The minimization of $E$ leads to the equations:
\begin{eqnarray}
1+\frac{\tilde g}{2(2\pi)^{3/2}L_z}\left [ 
1-\frac{2\pi}{3}\beta F\left ( \frac{L_\rho}{L_z} \right )
\right ]=0, 
\label{St1}
\\
L_z=\frac{1}{L_z^3}+\frac{\tilde g}{2(2\pi)^{3/2}L_\rho^2L_z^2}\left [ 
1-\frac{4\pi}{3}\beta G\left ( \frac{L_\rho}{L_z} \right )
\right ]
\label{St2}
\end{eqnarray}
where $F(\kappa)=(1-\kappa^2)^{-2} \left [-4\kappa^4-7\kappa^2+2+9\kappa^4 H(\kappa)\right ]$, 
$G(\kappa)=(1-\kappa^2)^{-2} \left [-2\kappa^4-+10\kappa^2+1-9\kappa^2 H(\kappa)\right ]$, and 
$\beta=\tilde g_d/\tilde g$. Eqs.~(\ref{St1}) and (\ref{St2}) admit a solution, and hence 
a localized wave, only under certain conditions. A simplified picture 
may be achieved by considering the fully two-dimensional situation in which the confinement 
in $z$ is strong enough to guarantee $L_z=1$. 
In that case, both kinetic and interaction energy scale as $1/L_\rho^2$. 
In absence of dipole-dipole interactions ($\tilde g_d=0$), and irrespective of the 
value of $L_\rho$, $E(L_\rho)$ is always either growing with $L_\rho$ (collapse instability) 
or decreasing with $L_\rho$ (expansion instability). This reflects the well-known fact that 
2D solitons are not stable in NLSE with contact interactions. 
In the case of a dipolar BEC, the situation is 
remarkably different, since the function $f$ depends explicitly on $L_\rho$. 
This allows for the appearance of a minimum in $E(L_\rho)$ (inset Fig.~\ref{fig:1}), which 
from the asymptotic values of $f$ ($f(0)=-1$ and $f(\kappa\rightarrow\infty)=2$) 
should occur if: 
\begin{equation}
\frac{\tilde g_d}{3\sqrt{2\pi}}<1+
\frac{\tilde g}{2(2\pi)^{3/2}}<\frac{-2\tilde g_d}{3\sqrt{2\pi}}.
\label{condition}
\end{equation}
A simple inspection shows that this condition can be fulfilled only if $\tilde g_d<0$, i.e. 
only if the dipole is tuned as previously discussed with $\phi>54.7^\circ$ (this is true 
also for $L_z\neq 1$). In that case, the tuning of the dipole-dipole interaction may allow for the 
observation of a stable 2D solitary wave, characterized by an internal energy 
$E_S<0$ (inset Fig.~\ref{fig:1}). Note that if $Na/l_z\gg 1$, then we arrive to the condition 
$|\beta|>3/8\pi\simeq 0.12$. 
We have compared this value with the result obtained 
from the direct resolution of Eq.~(\ref{GPE2D}) (see below) 
for the case of large $\tilde g$, obtaining 
stable 2D solitary waves for $|\beta|>0.12$, in excellent agreement with the Gaussian Ansatz.


In order to gain more understanding on the stability of 
the 2D solitary waves, we have analyzed the lowest 
lying modes of these structures, namely the breathing and quadrupolar modes. 
To this aim, we employ a Gaussian Ansatz of the form \cite{PerezGarcia,You2001}:
\begin{equation}
\Psi(\vec r;t)=\Psi_0\left ( \frac{x}{b_x},\frac{y}{b_y},\frac{z}{b_z}\right )
e^{
i\beta_x x^2+i\beta_y y^2+i\beta_z y^2}
\end{equation}
where $\{b_i(t),\beta_i(t)\}$, $i=x,y,z$, are time-dependent parameters,  
and insert this Ansatz into the corresponding Lagrangian density
\begin{eqnarray}
{\cal L}&=&\frac{i\hbar}{2}(\Psi\dot\Psi^*-\dot\Psi\Psi^*)+\frac{\hbar^2}{2m}|\nabla\Psi|^2+
\frac{1}{2}m\omega_z^2z^2|\Psi(\vec r,t)|^2+ \nonumber \\
&&\!\! \!\!\!\! \!\! 
\frac{g_c}{2}|\Psi(\vec r,t)|^4+\frac{1}{2}|\Psi(\vec r,t)|^2\int d\vec r'
V_d(\vec r-\vec r')|\Psi(\vec r',t)|^2.
\end{eqnarray}
After integrating $L=\int d\vec r{\cal L}$, we obtain the corresponding 
Euler-Lagrange equations for $\{b_i(t),\beta_i(t)\}$. Linearizing these equations around the 
stationary solution obtained from Eqs.~(\ref{St1}) and (\ref{St2}), we obtain the 
expressions for the frequencies of the lowest-lying modes \cite{You2001}.  
A typical behavior of the frequencies of the lowest modes is depicted in Fig.~\ref{fig:1}. 
The lowest-lying mode has for any value of $\beta$ a breathing character. For sufficiently small values of 
$|\beta|$, the breathing mode tends to zero, and eventually the system becomes unstable against expansion. This corresponds 
to the disappearance of the minimum in the inset of Fig.~\ref{fig:1}. In this regime, the 2D picture provides a good description of the 
physics of the problem, as shown in Fig.~\ref{fig:1}. For sufficiently large values of $|\beta|$, the 3D character of the system 
becomes crucial, leading to a different sort of instability, in this case against 3D collapse. 
This is reflected in the decrease of the frequency of the breathing mode.

As shown in Fig.~\ref{fig:1} a 2D calculation offers a good description of the problem for sufficiently 
small values of $|\beta|$. In that case the wavefunction factorizes as   
$\Psi(\vec r)=\psi(\vec\rho)\varphi_0(z)$, where $\varphi_0$ is the 
the ground-state of the harmonic oscillator in the $z$-direction. 
Employing this factorization, the convolution theorem, and the 
Fourier transform of the dipole potential, $\tilde V_d(k)=(4\pi/3)(3 k_z^2/k^2-1)$,
we arrive at the 2D NLSE:
\begin{eqnarray}
i\hbar\frac{\partial}{\partial t}\psi(\vec\rho,t)=
\left [ 
-\frac{\hbar^2}{2m}\nabla_\rho^2
+\frac{g}{\sqrt{2\pi}\l_z^2}|\psi(\vec\rho,t)|^2 + \right\delimiter 0 
\hspace{1cm}\nonumber \\
\left\delimiter 0  \frac{4\sqrt{\pi}g_d}{3\sqrt{2}l_z} 
\int \frac{d\vec k_\rho}{(2\pi)^2} 
e^{i\vec k_\rho\cdot\vec\rho} \tilde n(\vec k_\rho) h_{2D} \left (
\frac{k_\rho l_z}{\sqrt{2}}
\right )
\right ]
\psi(\vec\rho,t),
\label{GPE2D}
\end{eqnarray}
where $\tilde n$ is the Fourier transform of $n(\vec\rho)=|\psi(\vec\rho)|^2$, and 
$h_{2D}(k)=-2-3\sqrt{\pi}ke^{k^2}{\rm erfc}(k)$, with ${\rm erfc}(x)$ the 
complementary error function. Below, we employ Eq.~(\ref{GPE2D}) 
to analyze numerically the dynamics of 2D solitary waves.

\begin{figure}[ht] 
\begin{center}
\includegraphics[width=5.0cm]{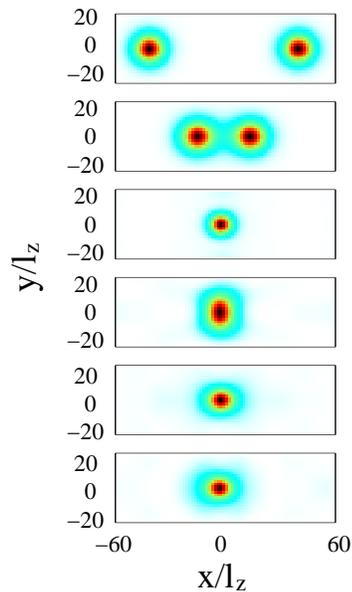}
\end{center} 
\vspace*{-0.2cm} 
\caption{Density plot of the 
fusion of two dipolar 2D solitary waves for $\tilde g=20$, $\beta=-0.5$, and 
$\tilde k=0.01$. From top to bottom $\omega_z t/2=0,1000,2000,3000,4000,5000$.}
\label{fig:2}  
\vspace*{-0.1cm}  
\end{figure}

Up to now, we have analyzed a single localized wave, showing that a stable solitary wave 
may exist under appropriate conditions. In order to deepen our understanding of these 
2D solutions, and their comparison with the solitonic solutions of the 1D NLSE, we have 
analyzed the scattering of two of these localized waves for different values of their 
initial center-of-mass kinetic energy, $E_{kin}$. Direct numerical simulations of the 
2D nonlocal NLSE show that the scattering of dipolar 2D solitary waves is inelastic. 
In particular, as shown in Fig.~\ref{fig:2}, for sufficiently slow localized waves 
(for the case considered in Fig.~\ref{fig:2}, $E_{kin}\leq 2.9 |E_S|$) 
two solitary waves merge when colliding. As observed in Fig.~\ref{fig:2}, the solitary 
waves, when approaching,  
transfer their center-of-mass kinetic energy into internal energy, transforming into 
a single localized structure. This structure, although localized, is in an excited state, 
and clear oscillations may be observed. For larger initial kinetic energies, the 
waves move apart from each other after the collision, but the transfer of kinetic energy 
into internal energy is enough to unbind the solitary wave, and the solitary waves are destroyed.
This inelastic character of the scattering of dipolar 2D solitary waves, 
clearly differs these solutions from the solitonic solutions of the 1D NLSE \cite{Zakharov}, and 
will be the subject of further investigation.

In the final part of this Letter, we would like to discuss some issues concerning the 
experimental realization of stable 2D solitary waves in ultracold dipolar gases. 
As we have previously mentioned, the experimental generation of these structures  
demands the tuning of the dipole-dipole interaction, and in particular the inversion of its sign. 
In addition to this condition, the dipole-dipole interaction must be sufficiently large, 
$|g_d|/g>0.12$. Unfortunately, for the case of Chromium, $|g_d|/g\simeq 0.03$ 
for a tuning angle $\phi=\pi/2$, and hence 
the tuning of the dipole-dipole interaction must be 
combined with a reduction of the contact interactions via Feshbach resonances. 
This combination results problematic due to technical limitations \cite{Pfau_Privat}. 
Therefore, the best candidate for the generation of 2D solitary waves is a 
condensate of polar molecules. As commented in the introduction of this Letter 
current developments open very optimistic perspectives for the achievement 
of a BEC of polar molecules in the very next future. 
Since appropriate molecules in the lowest vibronic state can present 
very large electric dipole moments \cite{DeMille}, and the dipole-dipole interaction may be  
tuned in these gases by means of rotating electric fields, a broad regime of values of 
$\beta$ may be available, and hence a stable 2D solitary wave should be observable.
\begin{figure}[ht] 
\begin{center}
\includegraphics[width=6.5cm]{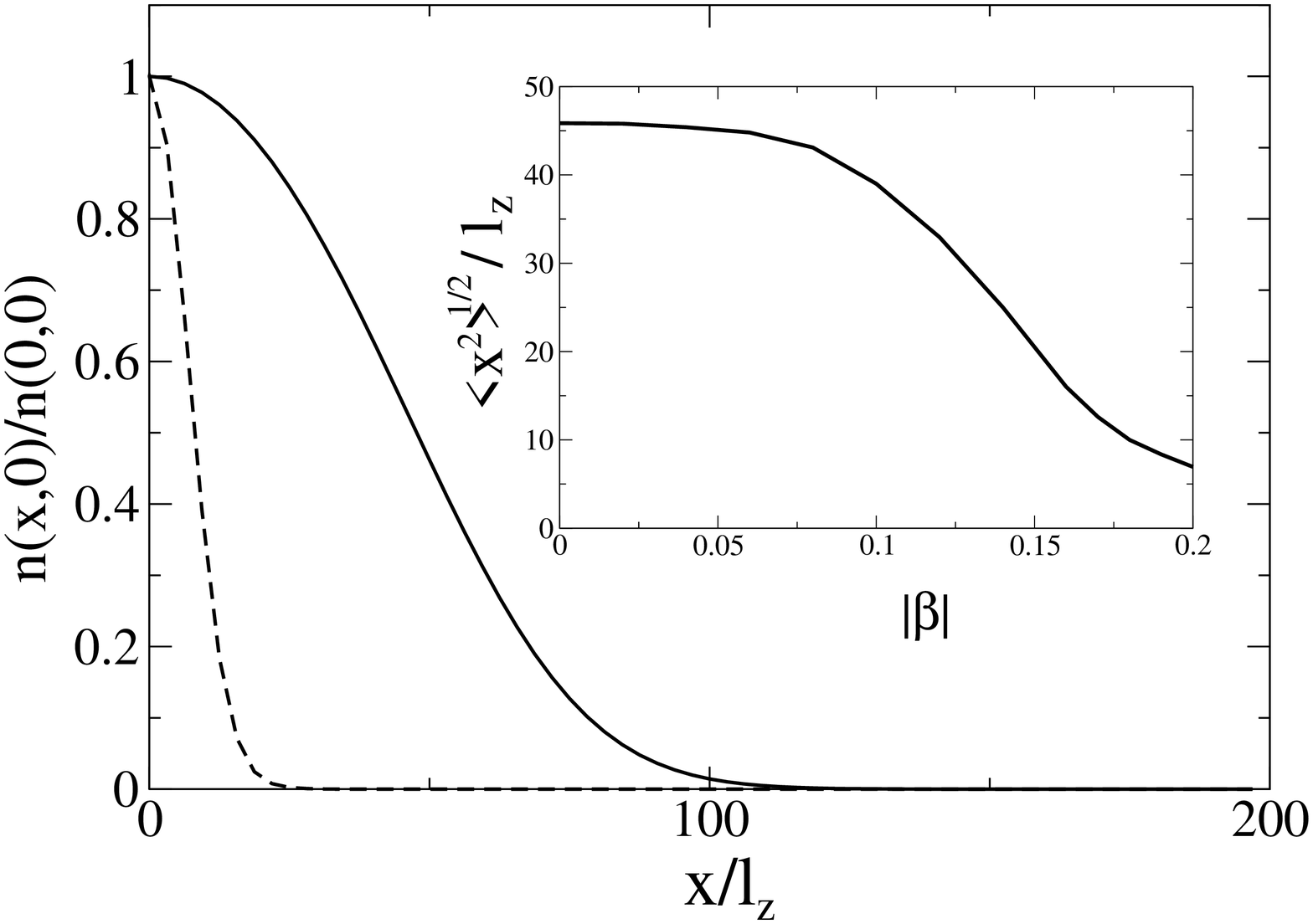}
\end{center} 
\vspace*{-0.2cm}  
\caption{Density profile for $\tilde g=200$, and $\beta=-0.10$ (solid) and $-0.20$ (dashed).
For both cases a $xy$ trap with oscillator length $l_\perp=40l_z$ has been employed. 
In the inset, width of the BEC as a function of $|\beta|$ for the same parameters.}
\label{fig:3}
\vspace*{-0.1cm}  
\end{figure}

In the previous discussion we did not consider any trapping on the $xy$ plane. 
In Fig.~\ref{fig:3} we consider 
the case of a loose $xy$-trapping. A large shrinking of the BEC density profile evidences 
the appearance of the solitary-wave solution. We would like to stress, that this large 
macroscopic effect may be scanned using the tuning techniques in polar molecules, 
it does not involve any collapse of the wavefunction (which is prevented by the dipole-dipole 
interactions), and it occurs for values of $|\beta|\ll 1$ 
for which the dipole-dipole interaction could be considered a perturbation to the 
short-range one. Finally, we should point out that analysis of the unperturbed propagation of 
the solitary wave may be performed by inducing bouncings within loose $xy$-traps, 
or by adiabatically removing the $xy$ confinement. 
Collisions of solitary waves could be performed by generating the waves in two separated 
dipolar traps, and subsequently releasing them.

In summary, we have shown that contrary to the case of short-range interacting BECs, 
stable 2D solitary waves can be generated in dipolar BEC (especially in molecular BEC) 
by means of tuning techniques, if the dipole-dipole interaction is sufficiently large. 
These solitary waves scatter inelastically, and may undergo fusion 
for sufficiently low scattering energies. For trapped gases, a significant 
shrinking of the BEC is predicted even for $|g_d|\ll g$.

Fruitful conversations with Y. Kivshar, D. Frantzeskakis, M. Fatori and T. Pfau are 
acknowledged. This work was supported by the Alexander von Humboldt Foundation.


\begin{thebibliography}{99}

\bibitem{Review} L. P. Pitaevskii and S. Stringari, {\it 
Bose-Einstein Condensation}, Oxford University Press (2003).

\bibitem{Chromium} A. Griesmaier et al., submitted.

\bibitem{Molec1} J.D. Weinstein et al., Nature (London) {\bf 395}, 148 (1998); 
H. L. Bethlem et al., Nature (London) {\bf 406}, 491 (2000); 
M. S. Ellio , J. J. Valentini, and D. W. Chandler, Science {\bf 302}, 1940 (2003); 
S. A. Rangwala et al., Phys. Rev. A {\bf 67}, 043406 (2002).

\bibitem{PA} A. J. Kerman et al., Phys. Rev. Lett. {\bf 92}, 153001 (2004); 
M. W. Mancini et al., Phys. Rev. Lett. {\bf 92}, 133203 (2004); 
C. Haimberger et al., Phys. Rev. A. {\bf 70}, 021402(R) (2004); 
D. Wang et al., Phys. Rev. Lett. {\bf 93}, 243005 (2004).

\bibitem{Feshbach} S. Inouye et al., Phys. Rev. Lett. {\bf 93}, 183201 (2004); 
C. A. Stan et al., Phys. Rev. Lett. {\bf 93}, 143001 (2004).

\bibitem{DeMille} J. M. Sage et al., physics/0501008.

\bibitem{Stability} S. Yi and L. You, Phys. Rev. A 61, 041604 (2000); 
K. G\'oral, K. Rz\c a\.zewski, and T. Pfau, Phys. Rev. A {\bf 61}, 051601 (2000); 
L. Santos et al., Phys. Rev. Lett. {\bf 85}, 1791 (2000).

\bibitem{Excitations} S. Yi and L. You, Phys. Rev. A {\bf 66}, 013607 (2002);
K. G\'oral and L. Santos,  Phys. Rev. A {\bf 66}, 023613 (2002).

\bibitem{Roton} L. Santos, G. V. Shlyapnikov, and M. Lewenstein, 
Phys. Rev. Lett. {\bf 90}, 250403 (2003); S. Giovanazzi, and D. H. J. O'Dell, 
Eur. Phys. J. D {\bf 31}, 439 (2004).

\bibitem{DipLat} K. G\'oral, L. Santos, and M. Lewenstein, Phys. Rev. Lett. {\bf 88}, 170406 (2002).

\bibitem{FQHE} M. A. Baranov, Klaus Osterloh, and M. Lewenstein, Phys. Rev. Lett. {\bf 94}, 070404 (2005).

\bibitem{QInf} D. DeMille, Phys. Rev. Lett. {\bf 88}, 067901 (2002).

\bibitem{Chemistry} E. Bodo, F. A. Gianturco, and A. Dalgarno, J. Chem. Phys. {\bf 116}, 9222 (2002).

\bibitem{4WM} L. Deng {\it et al.}, Nature {\bf 398}, 218 (1999). 

\bibitem{Collapse} E. A. Donley {\it et al.}, Nature {\bf 412}, 295 (2001).

\bibitem{Solitons} J. Denschlag {\it et al.}, Science {\bf 287}, 97 (2000);
S. Burger {\it et al.}, Phys. Rev. Lett. {\bf 83}, 5198 (1999); 
L. Khaykovich {\it et al.}, Science {\bf 296}, 1290 (2002); 
K. E. Strecker {\it et al.}, Nature {\bf 417}, 150 (2002); B. Eiermann {\it et al.}, 
Phys. Rev. Lett. {\bf 92}, 230401 (2004) 

\bibitem{Zakharov} 
V. E. Zakharov and A. B. Shabat, Zh. Eksp. Teor. Fiz. {\bf 61}, 118 (1971)
[Sov. Phys. JETP {\bf 34}, 62 (1972)]; V. E. Zakharov and A. B. Shabat,
Zh. Eksp. Teor. Fiz. {\bf 64}, 1627 (1973) [Sov. Phys. JETP {\bf 37}, 823 (1973)].

\bibitem{Kivsharpapers} V. A. Mironov, A. M. Sergeev, and E. M. Sher,  
Sov. Phys. Dokl. {\bf 26}, 861 (1981); 
W. Kr\'olikowski and O. Bang, Phys. Rev. E {\bf 63}, 016610 (2000); 
N. N. Rosanov {\it et al.}, Phys. Lett. A {\bf 293}, 45 (2002); 
N. N. Rosanov et al., JETP Lett. {\bf 77}, 89 (2003); 
N. I. Nikolov {\it et al.}, Opt. Lett. {\bf 29}, 286 (2004);
A. I. Yakimenko, Y. A. Zaliznyak and Y. Kivshar, nlin.PS/0411024.

\bibitem{Nematics} M. Peccianti {\it et al.}, Nature {\bf 432}, 733 (2004). 

\bibitem{Bang} O. Bang, W. Krolikowski, J. Wyller and J. J. Rasmussen, 
Phys. Rev. E, {\bf 66}, 046619 (2002).

\bibitem{Tuning} S. Giovanazzi, A. G\"orlitz, and T. Pfau, Phys. Rev. Lett. {\bf 89}, 130401 (2002).

\bibitem{PerezGarcia} V. M. P\'erez-Garc\'\i a, {\it et al.}, 
Phys. Rev. Lett. {\bf 77}, 5320 (1996).

\bibitem{You2001} S. Yi and L. You, Phys. Rev. A {\bf 63}, 053607 (2001).

\bibitem{Pfau_Privat} T. Pfau, private communication.

\end{thebibliography}
\end{document}